\documentclass[twocolumn,prb,floatfix,superscriptaddress,nobibnotes]{revtex4-2}
\usepackage{graphicx}
\usepackage{enumerate}
\usepackage[colorlinks=true,citecolor=blue]{hyperref}
\usepackage{textcomp}
\usepackage{amsmath}
\usepackage{amssymb}
\usepackage{pgf}
\usepackage{subfigure}
\usepackage[english]{babel}
\usepackage[normalem]{ulem}
\usepackage{tabularray}
\usepackage{array}

\def\ud{\mathrm{d}}

\renewcommand{\d}{{\bf d}}

\renewcommand{\l}{{\bf l}}

\graphicspath{{./}{./}}

\begin{document}
\title{Multiterminal transport spectroscopy of subgap states in Coulomb-blockaded superconductors}

\author{Rub\'en Seoane Souto}
\affiliation{Center for Quantum Devices, Niels Bohr Institute, University of Copenhagen, 2100 Copenhagen, Denmark}
\affiliation{Division of Solid State Physics and NanoLund, Lund University, S-22100 Lund, Sweden}

\author{Matteo M. Wauters}
\affiliation{Center for Quantum Devices, Niels Bohr Institute, University of Copenhagen, 2100 Copenhagen, Denmark}
\affiliation{Niels Bohr International Academy, Niels Bohr Institute, University of Copenhagen, 2100 Copenhagen, Denmark}

\author{Karsten Flensberg}
\affiliation{Center for Quantum Devices, Niels Bohr Institute, University of Copenhagen, 2100 Copenhagen, Denmark}

\author{Martin Leijnse}
\affiliation{Center for Quantum Devices, Niels Bohr Institute, University of Copenhagen, 2100 Copenhagen, Denmark}
\affiliation{Division of Solid State Physics and NanoLund, Lund University, S-22100 Lund, Sweden}

\author{Michele Burrello}
\affiliation{Center for Quantum Devices, Niels Bohr Institute, University of Copenhagen, 2100 Copenhagen, Denmark}
\affiliation{Niels Bohr International Academy, Niels Bohr Institute, University of Copenhagen, 2100 Copenhagen, Denmark}

\begin{abstract}
Subgap states are responsible for the low-bias transport features of hybrid superconducting--semiconducting devices. Here, we analyze the local and nonlocal differential conductance of Coulomb-blockaded multiterminal superconducting islands  that host subgap states with different spatial structures. The emerging patterns of their transport spectroscopy are used to characterize the possible topological nature of these devices and offer the possibility of controlling their transport properties.
We develop a next-to-leading order master equation to describe the multiterminal transport in superconductors with both strong Coulomb interactions and multiple subgap states, coupled with metallic leads. 
We show that the nonlocal differential conductance characterizes the spatial extension of the subgap states and signals the presence of degenerate bound states with a finite support on different parts of the device.
Additionally, it displays sharp sign changes as a function of the induced charge of the superconductor, signaling energy crossings among its lowest excited states.
\end{abstract}

\maketitle

\section{Introduction}
Hybrid systems fabricated with superconducting and semiconducting materials are considered a key ingredient for the development of quantum technologies and have demonstrated a potential for quantum information storage and processing \cite{Janvier_Science2015,Hays_PRL2018,Hays_Science2021,Prada_review}. In particular, the advent of topological superconductors has raised the hope of realizing protected quantum devices, robust to decoherence induced by local perturbations~\cite{NayakReview}. 

In one dimension, topological superconductors host Majorana subgap modes \cite{Kitaev_2001}, predicted to be nonabelian anyons and to give rise to several nonlocal properties \cite{Alicea_2012,LeijnseReview,BeenakkerReview_20,AguadoReview}. These Majorana modes are exponentially localized at the system edges but, importantly, each pair of them defines a nonlocal quasiparticle state, with vanishing energy in the ideal case. In this context, distinguishing between local and nonlocal subgap states becomes relevant for identifying devices that are potentially in the topological regime.

Subgap states dominate the transport properties of superconducting devices at low voltage bias and they can be experimentally detected with tunneling spectroscopy.
In particular, zero-energy Majorana modes in grounded topological superconductors reveal themselves through a quantized conductance peak at zero bias \cite{Sengupta_PRB2001,Law2009,Flensberg_PRB2010, Wimmer_2011}. However, also trivial states can mimic similar transport features in local spectroscopy, see for example Refs. \cite{Pikulin2012,Prada_PRB2012,Kells_PRB12,Liu2017,Moore_PRB18,Vuik_SciPost19,Awoga_PRL2019,Pan_PRR20,Avila_ComPhys2019,Cayao_PRB2021}. This motivated further analyses of the nonlocal conductance in multiterminal setups, where the device couples to a grounded superconductor \cite{Rosdahl_PRB2018,Danon_PRL2020,Hess_PRB2021,Melo_SciPost2021,Pikulin_arXiv2021,Singh_arXiv2022,Maiani_arXiv2022,Thanos_arXiv2022}, with experiments performed in nanowires \cite{Gramich_PRB2017,Menard_PRL2020,Puglia_PRB2021,Martinez_arXiv2021,Guanzhong_arXiv2022,Dvir_arXiv2022}, and devices based on two-dimensional electron gases \cite{Poschl_arXiv2022,Poschl_arXiv2022_2,Banerjee_arXiv2022,Aghae_arXiv2022}.

When the superconductor is not grounded but exhibits a strong charging energy, the intricate interplay between superconductivity and many-body interactions yields a rich phenomenology.
In the topological regime, Coulomb blockade effects provide effective tools to initialize, read out, and measure coherence times in topological qubits, thus constituting a key element to design platforms for quantum information devices~\cite{Heck_NPJ2012, Terhal_PRL2012, Hyart_PRB2013, Plugge_PRB2016, Aasen_PRX2016, Landau_PRL2016, Plugge_NJP2017, Karzig_PRB2017,Manousakis2020, Nitsch_arXiv2022, Souto_SciPost2022}. 

Transport properties of floating superconducting islands have been extensively studied experimentally in semiconducting nanowires, with the goal of extracting quasiparticle poisoning times \cite{Higginbotham_NatPhys2015,Albrecht_PRL2017}, showing the doubling of Coulomb peaks in the high-field regime~\cite{Albrecht_Nature2016,Farrell_PRL2018,Hansen_PRB2018,Vaitiekenas_Science2020}, and investigating the exponential protection \cite{Albrecht_Nature2016} and spin polarization \cite{Vaitiekenas_PRB2022} of subgap states. These systems have been explored using interferometry \cite{Whiticar_NatCom2020} and reflectometry techniques \cite{Razmadze_PRAp2019,Sabonis_APL2019}. 
However, most of the studies so far, both experimental and theoretical, focused on local conductance measurements of two-terminal devices; nonlocal transport properties of multiterminal Coulomb-blockaded devices remain largely unexplored. 

A more complex typology of devices is based on two nanowires coupled via a floating superconductor, sharing a common charging energy. In the case where the device couples to more than two leads, the topological Kondo effect is predicted to appear, and its realization would be an unequivocal signature of the Majorana nonlocality~\cite{Beri_PRL2012,Altland_PRL2014,Zazunov_NJP2014}. As such, systems with Coulomb-blockaded double nanowires are at the focus of recent experiments~\cite{kanne2022double,Vekris_DNW2022} and theoretical investigations \cite{Ekstrom_PRB2020}. 

In this work, we propose multiterminal quantum transport measurements as a way to distinguish between local and nonlocal subgap states in Coulomb-blockaded superconducting devices. 
 We focus on two geometries, namely a single and a double superconducting nanowire device.
We employ a second-order master equation approach that allows for studying both sequential and cotunneling transport in superconducting devices with strong Coulomb interaction, multiple subgap states, and the coupling to two or more metallic leads.
In particular, we show that sequential transport at low bias voltages is blocked whenever the lowest-energy subgap state does not couple two of the leads.  
Indeed, even though the Coulomb interaction is highly nonlocal, transport is dominated by the local projection of subgap states, providing indications about their spatial structure.
Moreover, a three-terminal Coulomb-blockaded device displays abrupt changes in the current direction at fixed bias potential as a function of the charge $n_g$ induced on the superconducting island, due to energy crossings in the excitation spectrum, similarly to non-superconducting Coulomb-blockaded quantum dots \cite{Zak_PRB2008}.
Concerning the double nanowire  geometry, we investigate the local and nonlocal conductance associated to different spatial structures of the subgap states, aiming to facilitate the interpretation of future experiments.

The rest of the paper is organized as follows.
In Sec.~\ref{sec:model} we present the effective zero bandwidth model we use to describe the low-energy behavior of the devices and the formalism used to compute the sequential and cotunneling current signals.
In Sec.~\ref{sec:SNW} we investigate both the two-terminal and three-terminal conductance of a single nanowire with one or multiple subgap states.
In Sec.~\ref{sec:DNW} we analyze the transport signatures associated to different structures of the subgap states in the double nanowire geometry.
We draw our conclusions and discuss future developments in Sec.~\ref{sec:conclusion}.
The appendices contain more supporting results and a detailed derivation of the tunneling rates calculation.

\begin{figure}[h] \centering
\includegraphics[width=1\linewidth]{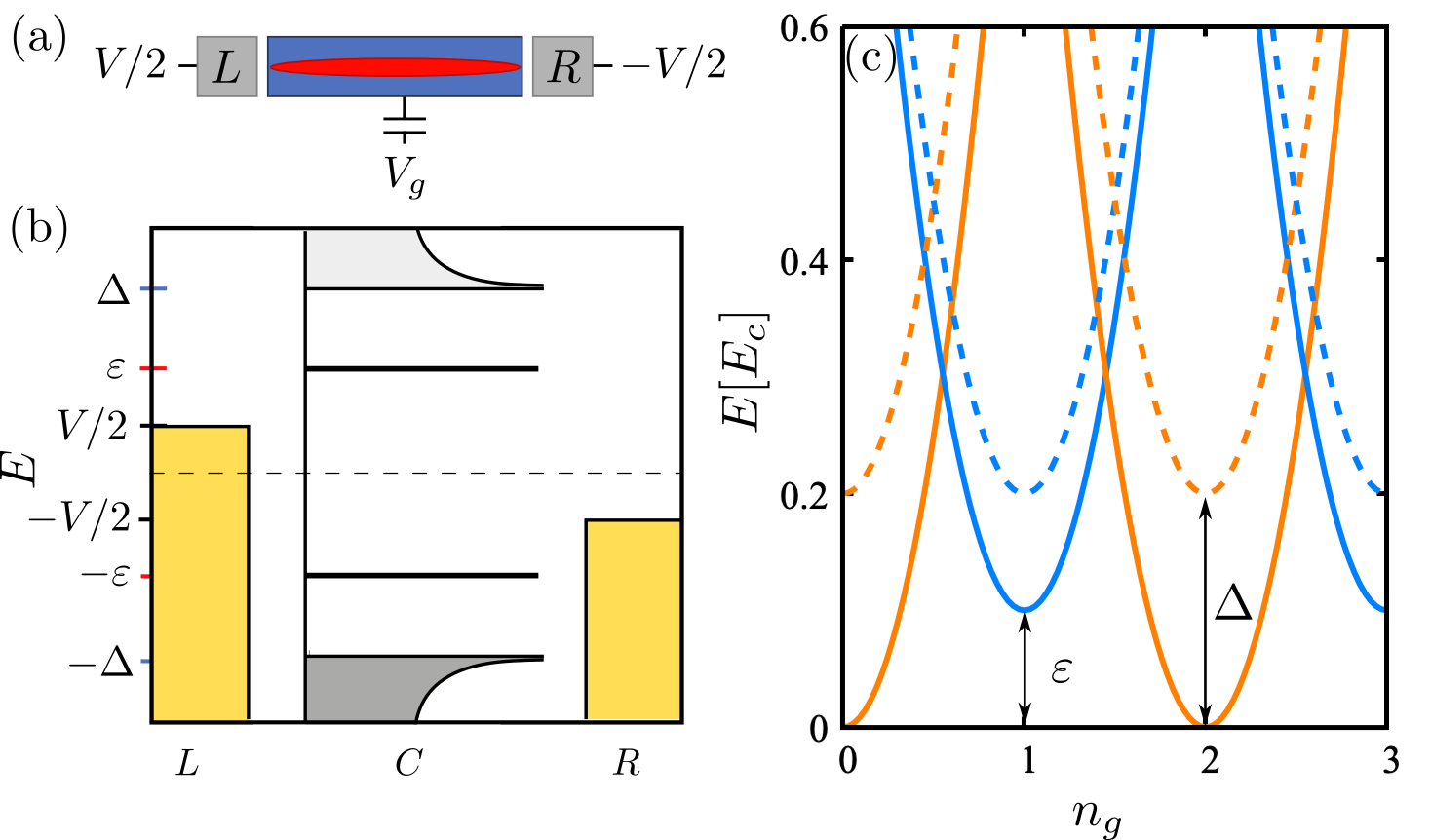}
\caption{(a) Sketch of a Coulomb-blockaded hybrid superconducting nanowire device, hosting a subgap state (red). In this example, the nanowire couples to two leads (indicated with $L$ and $R$). (b) Single-particle energies of the island, containing a subgap state at energy $\varepsilon$ (black line), quasiparticle excitations above the gap $\Delta$, and the lead states displaying Fermi-Dirac distributions (yellow) at different chemical potentials. (c) Lowest eigenergies in the presence of interactions as a function of the offset charge, $n_g=\alpha V_g$ with $\alpha$ being the lever arm. Blue (orange) lines correspond to states with odd (even) number of electrons in the island. We show an example where the nanowire hosts a spinless subgap state, where the solid and dashed lines correspond to the ground state and the excited state in each charge sector. 
}\label{Fig0}
\end{figure}

\section{Model and methods}\label{sec:model}
We consider a superconducting island coupled to normal metallic electrodes, given by the Hamiltonian
\begin{equation}
    H=H_{\rm L}+H_{\rm I}+H_{\rm T}\,,
\end{equation}
where the leads are described by 
\begin{equation}
    H_{\rm L}=\sum_{\rm \nu,k,\sigma}\left( \xi_{\nu k\sigma}-\mu_\nu \right)c^{\dagger}_{\nu k\sigma}c_{\nu k\sigma}\,,
\end{equation}
with the electron energy, $\xi_{\nu k\sigma}$, and the annihilation operator, $c_{\nu k\sigma}$, referring to an electron in lead $\nu$ with momentum $k$ and spin $\sigma \in \{\uparrow, \downarrow\}$. We assume that each lead remains in internal equilibrium, described by a Fermi-Dirac distribution $n_{\rm{F}}$ with chemical potential $\mu_\nu$ and temperature $T$. 

The superconducting island is described by
\begin{equation}
    H_{\rm I}=\sum_{j,\sigma}\varepsilon_{j\sigma} \gamma_{j\sigma}^\dagger \gamma_{j\sigma}+E_{\rm el}(N)\,,
    \label{H_island}
\end{equation}
where $j$ labels a state with energy $\varepsilon_{j\sigma}$, and the Bogoliubov-de Gennes (BdG) operators read
\begin{equation}
    \gamma_{j\sigma}=u_j d_{j\sigma}-\rho_\sigma \, v_je^{-i\phi}d_{j\bar{\sigma}}^\dagger\,.
\end{equation}
Here, $d$ and $d^\dag$ are annihilation and creation operators of electrons in the island, $u_j$ and $v_j$ are the BdG coefficients for the subgap states, $\phi$ is the superconducting phase operator, such that $e^{-i\phi}$ annihilates a Cooper pair in the island, and $\rho_\sigma=\pm$ for $\sigma=\uparrow,\downarrow$. 

The electrostatic repulsion term is given by
\begin{equation}
    E_{\rm el}(N)=E_{\rm C}(N-n_{\rm g})^2\,,
\end{equation}
where $n_{\rm g}$ is the dimensionless gate-induced charge offset and $N$ the (excess) electron number operator, accounting for the fermionic occupation of the states and the number of Cooper pairs on the island.
Figure~\ref{Fig0}(a) sketches the device described by this model, in the case of only two leads and a single nanowire. Panels (b) and (c) represent the superconducting quasiparticle spectrum and the many-body energy levels,  respectively.

Finally, the tunneling between leads and superconducting island  is described by the Hamiltonian
\begin{equation}
    \begin{split}
    H_T & =\sum_{\nu, k,j,\sigma} \left(t_{\nu k\sigma}c_{\nu k\sigma}^\dagger d_{j\sigma}+ {\rm H.c.}\right)
	\end{split}
\end{equation}
where $t$ is the tunneling amplitude. The corresponding tunneling rates are defined as $\Gamma_{\nu,j,\sigma}=2\pi\rho_{F\nu} |t_{\nu, j,\sigma}|^2$, and we consider them $k$-independent (wideband limit), where $\rho_{F\nu}$ is the density of states at the Fermi level of the lead $\nu$.

In this work, we employ a zero-bandwidth approximation for the superconducting island, meaning that we consider only two relevant energy levels $\varepsilon_{0,\uparrow} = \varepsilon$ and $\varepsilon_{1,\sigma} = \Delta > \varepsilon$.
The first represents a low-energy spin-polarized subgap state whose spatial structure and degeneracy will be the main focus of our work.
The state at energy $\Delta$ is a spin-degenerate level and represents an effective description of the proximitized superconducting continuum which extends through the whole hybrid device.
This approximation models a short nanowire where the quasiparticle spectrum above the proximity induced superconducting gap displays a quantization determined by the finite system size. The corresponding lowest energy excitation can mediate elastic cotunelling processes through the device. 
A possible alternative is to consider a number of degenerate states at energy $\Delta$, each coupled to a single lead.
Each of these states models a metallic quasiparticle continuum with a strong relaxation towards the bottom of the continuum band. With this choice, the states at energy $\Delta$ can not mediate elastic cotunneling transport.
Throughout the rest of the paper, we focus on the first scenario, where each nanowire hosts a single extended superconducting continuum, to avoid an excessive increase of the computational run time.
However, this choice does not qualitatively influence the transport features at low bias, where the conductance is dominated by the subgap states.
See Fig.~\ref{S1} in App.~\ref{app:SR} for a comparison of the two cases.
\subsection{Formalism}\label{ssec:formalism}
In this work, we focus on the regime where 
$\Gamma_{\nu,j,\sigma}\ll T$. In this weak coupling limit, a perturbation theory in the tunneling amplitude provides accurate results.
Since we are not interested in the thermal excitations of the superconducting device, we additionally consider temperatures much smaller than the remaining energy scales.
To compute the tunneling rates, we use the $\mathcal{T}$-matrix formalism, describing the transition probability between two states
\begin{equation}
    \Gamma_{i\to f}=2\pi\left|\left\langle f\left| \mathcal{T} \right|i\right\rangle \right|^2 W_{if}\delta(E_i-E_f)\,,
	\label{rates_def}
\end{equation}
where $W_{if}$ weights the rate through thermal distributions of the electrons in the leads, $\delta$ is the Dirac delta functional ensuring energy conservation between the initial and final states ($E_i=E_f$) and 
\begin{equation}
    \mathcal{T}=H_{\rm T}+H_{\rm T}\frac{1}{E_i-H_{\rm L}-H_{\rm I}+i0^+}\mathcal{T}\,,
    \label{T_definition}
\end{equation}
which can be truncated at the desired order. In Eq.~\eqref{rates_def} and in the rest of the paper, we set $\hbar=1$. The linear term in Eq. \eqref{T_definition} describes the sequential tunneling; the higher order terms define, among other processes, the cotunneling contributions, which become progressively more important when the tunnel amplitudes $t$ increase. In Sec. \ref{Sec::tunnelingRates} of the appendix, we provide expressions for the sequential and the (second-order) cotunneling rates, used in this work to evaluate the transport through the superconducting islands.

The quantum state of the island is described by $\left|a\right\rangle$=$\left|N,N_{\rm C},\textbf{n}\right\rangle$, where $N_C$ is the number of Cooper pairs in the island and $\textbf{n}$ is a vector representing the occupation of the subgap and continuum excited states. The time derivative of the occupation probability of a given state is given by the master equation
\begin{equation}\label{eq:master}
	\dot{P}_a=\sum_b\left[-\Gamma_{a\to b}P_a+\Gamma_{b\to a}P_b\right]\,.
\end{equation}
Here, we assume that coherences between different states in the superconducting island are negligible; indeed, we consider either models without near degeneracies where coherences cannot develop or with no coupling between degenerate states, due to different parity or spatial support.
In the stationary limit, the system does not evolve in time and $\dot{P}^{\rm{stat}}_b=0$. These conditions, together with the normalization  $\sum_b P^{\rm{stat}}_b=1$, form a linear system of equations for the occupation probabilities of the island states. Using the resulting stationary distribution, the current flowing from lead $\nu$ to the device is determined by the island transition rate to other states as
\begin{eqnarray}\label{eq:current}
        I_\nu=\sum_{a,b}\left[s\Gamma^{\rm{seq}}_{\nu,b\to a}
	     +\sum_{\nu' \neq \nu} \delta n(\nu)\,\Gamma^{\rm{cot}}_{\nu,\nu',b\to a} \vphantom{\frac{1}{2}}\right]P^{\rm{stat}}_b\,,
\end{eqnarray}
where we take $s=+1$ ($s=-1$) for electrons tunneling in (out) of the island and we set the electron charge to unity. Here, $\delta n(\nu)$ is the net charge transferred from lead $\nu$ in a cotunneling process, and the sum runs over all the possible rates connecting the island state $\left|b\right\rangle$ with any state $\left|a\right\rangle$ (see Appendix \ref{app:cot} for more detail on the cotunneling rates $\Gamma^{\rm{cot}}$).

In Eqs.~\eqref{eq:master} and \eqref{eq:current} we are considering processes that change the island occupation by $\pm$1 electron, while neglecting local and crossed Andreev reflection processes~\cite{vanHeck_PRB2016}.
This approximation is justified when the charging energy is the dominant energy scale, in particular $E_c \gg \Delta$, meaning that there is a large energy penalty for changing the charge of the device by $2e$.
Throughout the rest of the paper, we will indeed consider this strongly Coulomb-blockaded regime.
\section{Single nanowire geometry}\label{sec:SNW}
In this section, we focus on resolving the spatial structure of subgap states in a single nanowire device coupled to two or three normal metallic leads. Regarding the spatial profile of the subgap state, we consider three situations: ({\em i}) a single subgap state delocalized at both ends of the nanowire but vanishing in its bulk, which we refer to as a ``Majorana-like scenario'', ({\em ii}) two degenerate subgap states each localized close to one end of the device, ({\em iii}) an extended subgap state that couples to all leads independently from their position. In all these cases, we consider that the subgap states are strongly spin-split, behaving as effectively spinless. This situation can be achieved by a strong magnetic field or by proximity with ferromagnetic materials, and corresponds to standard models to achieve the topological regime~\cite{Lutchyn_NatReview2018}.
Our goal is to distinguish between these possible scenarios based on the structure of the cotunneling steps and the nonlocal conductance.

\subsection{Two-terminal device}
\begin{figure}[h] \centering
\includegraphics[width=1\linewidth]{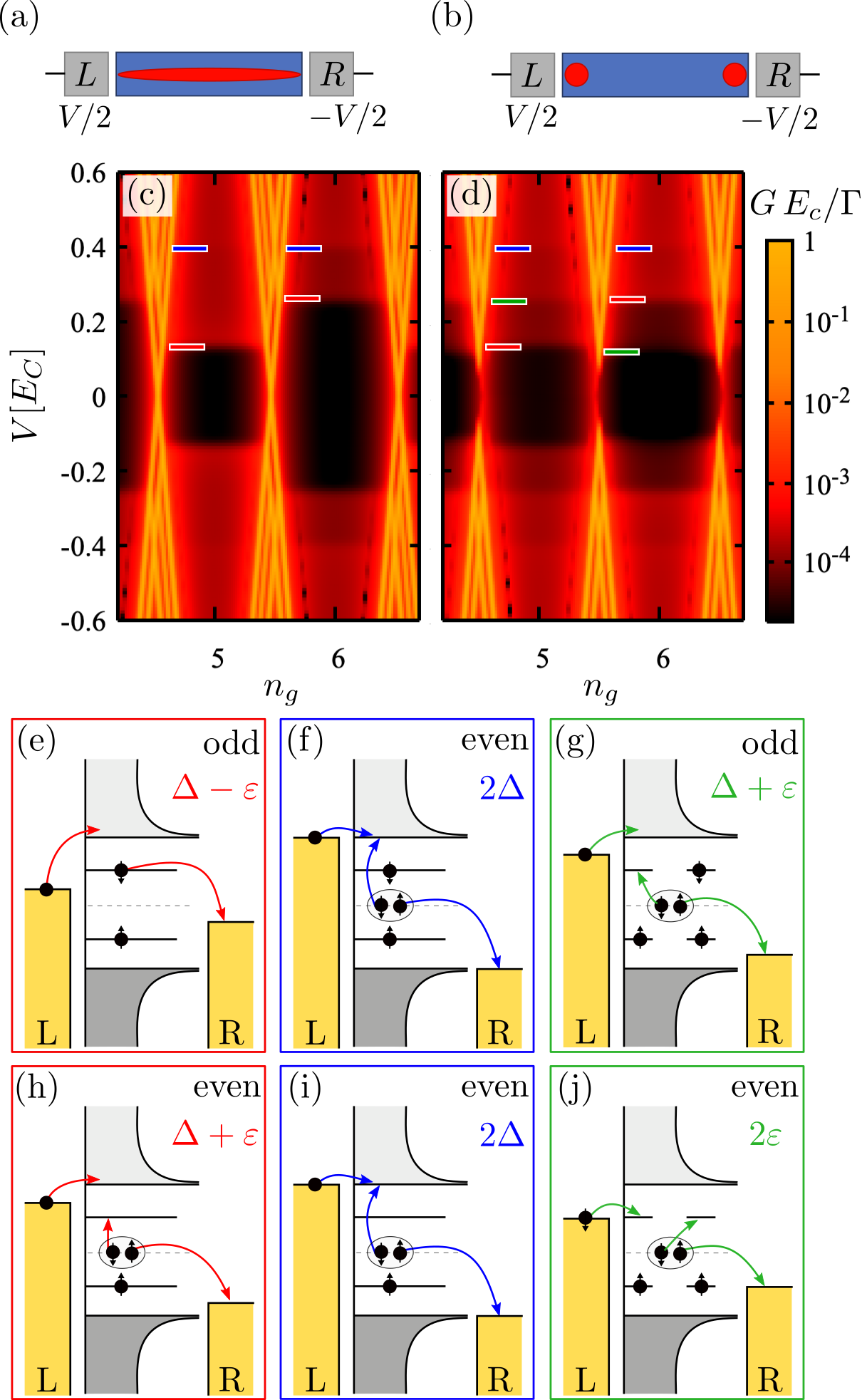}
\caption{Sketches of a nanowire coupled to two leads, hosting an extended (a) and two local states (b). Panels (c) and (d) show the differential conductance (in logscale) corresponding to cases (a) and (b), respectively, for subgap states with energies $\varepsilon=0.3\Delta$. We show results for $\Delta=0.2E_C$ and $\Gamma=5\cdot10^{-5}E_C$. The color tics denote the dominant inelastic cotunneling processes, appearing at $V=2\Delta$ (blue), $V=\Delta\pm\varepsilon$ for the even/odd Coulomb valley (red), and $V=2\varepsilon$, $\Delta+\varepsilon$ (green). Panels (e-j) show examples of the different cotunneling processes.
}\label{Fig1}
\end{figure}

We first focus on a two-terminal device, with leads coupled to the nanowire ends, which can host either a single subgap state or two degenerate ones located at the ends, see Figs. \ref{Fig1} (a) and (b).
For simplicity, we choose a symmetric voltage drop at the contacts between the device and $L$ and $R$ leads by setting their chemical to $\mu_L=+V/2$ and $\mu_R=-V/2$, where $V$ is the voltage bias.

If the subgap levels have zero energy $\varepsilon=0$, which is the case for ideal Majorana zero energy modes, the Coulomb blockade structure is qualitatively similar in the two cases (data not shown).
Instead, when $\varepsilon \neq 0$, the most evident difference is the strong reduction of the conductance peaks at $V=0$ in the case of local states.
This can be seen by comparing the conductance at $V\sim0$ and half-integer $n_g$ in panels (c) and (d) of Fig.~\ref{Fig1}. 
Indeed, when the subgap states do not couple to both leads, transport is suppressed at small biases because the subgap states cannot support resonant tunneling of electrons.
This results in a much weaker conductance for $|V| < 2\varepsilon,\ \Delta-\varepsilon$ with respect to the situation where the subgap state connects both ends of the nanowire. However, zero-bias tunneling is still possible through elastic cotunneling, with an amplitude $\sim \Gamma_L\Gamma_R$~\cite{vanHeck_PRB2016}.

We also observe a clear difference in the structure of the cotunneling steps, highlighted by the colored tics in Figs.~\ref{Fig1} (c) and (d). In the case of an extended state, two inelastic cotunneling steps appear at $V=\Delta\pm\varepsilon$ for the even and the odd valleys, corresponding to transitions to the lowest excited states (red tics) \cite{Vaitiekenas_PRB2022}. In the even valley, it corresponds to the splitting of a Cooper pair, whose electrons end up in the subgap and the continuum of states. In the odd valley, the lowest excitation corresponds to promoting the electron in the subgap state to the continuum. There are two additional steps due to the breaking of a Cooper pair into two electrons that enter the quasiparticle continuum (blue tics).

In the case with two states with local support, each coupled to a single lead, the number of cotunneling steps increases and, in particular, the threshold for inelastic cotunneling in the even-parity valleys decreases substantially. Indeed, the lowest excited state with even parity has an energy given by the sum of the energies of the two subgap states [see panel (j) in Fig. \ref{Fig1}]. In the case considered in the figure with the two states having the same energy, this threshold is $2\varepsilon$. We note that this process does not require the creation of a quasiparticle excitation above the superconducting gap, reducing the conductance step onset with respect to the nondegenerate situation.

\subsection{Three-terminal device}
\begin{figure}[h] \centering
\includegraphics[width=1\linewidth]{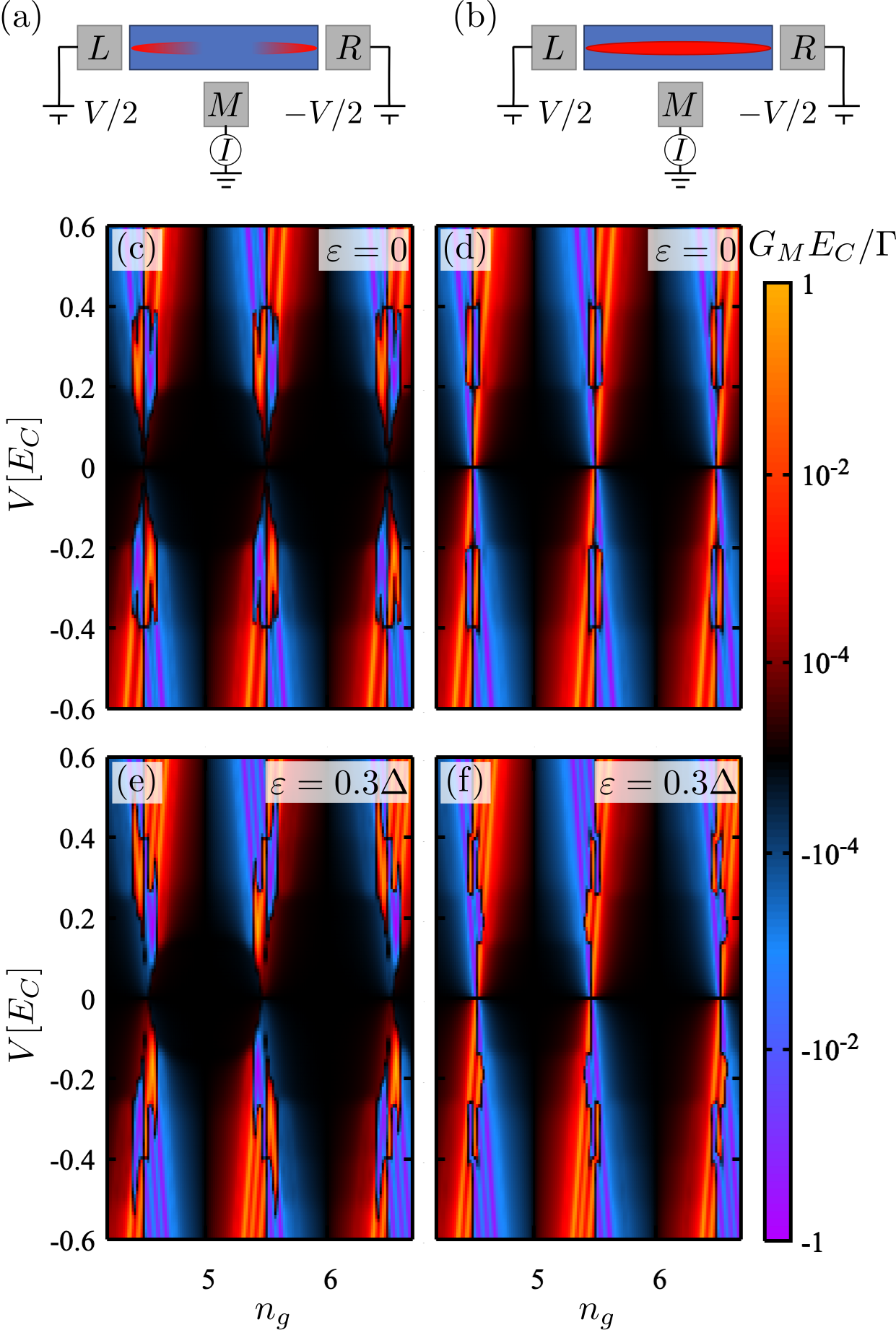}
\caption{Nonlocal differential conductance $G_M=\frac{\ud I_M}{\ud V}$ through a grounded lead coupled to the middle of the wire. (a) and (b) respectively depict situations where the middle lead does not and does couple to the subgap state (red blobs). Panels (c) and (d) show the nonlocal conductance for a case where the subgap state energy is $\varepsilon=0$, while panels (e) and (f) show cases for $\varepsilon=0.3\Delta$. 
Panels (c) and (e) refer to the subgap state structure shown in (a) while panels (d) and (f) correspond to (b).
The remaining parameters are the same as in Fig. \ref{Fig1} and all the tunnel couplings are considered to be equal.
}\label{Fig2_v2}
\end{figure}

Next, we analyse a three terminal device, where two leads ($L$ and $R$) couple to the nanowire ends and the third ($M$) probes the center of the device.
Therefore, the $M$ lead acts as a probe of the subgap support at the middle of the wire, distinguishing local and nonlocal states. 
The $L$ and $R$ leads are biased, with a chemical potential $\mu_L=+V/2$ and $\mu_R=-V/2$, respectively, while we consider that $M$ is grounded ($\mu_M=0$).
The goal is to use the nonlocal conductance $G_{M}=\frac{\ud I_M}{\ud V}$ to detect whether the subgap state extends inside the device or is localized only on the edges.
These two scenarios are sketched in Figs.~\ref{Fig2_v2} (a) and (b).
Furthermore, we explore the situations where the subgap states' energy $\varepsilon$ is zero, Figs.~\ref{Fig2_v2} (c)-(d), or finite, Figs.~\ref{Fig2_v2}(e)-(f).

While the local conductance (through $L$ and $R$) shows little dependence on the presence of a third lead and how it couples with the device (data not shown), the nonlocal conductance displays interesting features.
First, notice that $G_{M}$ is an odd function of the voltage bias $V$, thanks to the symmetry $R \leftrightarrow L$ we impose in our model: an inversion of the bias only changes the direction of the current flow between the $L$ and $R$ lead, while the current $I_M$ remains insensitive to the sign of $V$ since the lead is always grounded. Hence, the differential conductance ($G_{M}=\frac{\ud I_M}{\ud V}$) is an odd function of the voltage bias. 
When the lead does not couple to the subgap state, the conductance at $V=0$ is suppressed, as clearly seen by comparing left and right panels in Fig.~\ref{Fig2_v2}. Even though this effect is somehow trivial because low bias transport only involves the lowest energy level in a gapped system, its consequence is of great importance: the middle lead probes the density of states inside the wire. 
Hence, the $M$ lead can resolve the spatial profile of subgap states and detect whether they have a nonvanishing projection on a specific portion of the device.

The non-local conductance also shows a peculiar sign dependence on $n_g$, illustrated by the sharp $G_M$ jumps in Fig.~\ref{Fig2_v2}. 
These sign changes appear whenever there is a crossing between ground states or excited states with different parities, corresponding to the crossing of the parabolas in Fig. \ref{Fig0} (c) (see Appendix \ref{app:SR} for more details).
Their dependence on $V$, instead, arises from the necessity of having a chemical potential on the $L$ and $R$ leads large enough to populate the excited states coupled also with the $M$ lead.
This behavior can be better understood by considering the energy dependence of the superconducting island states on $n_g$ and the sequential tunneling rates (see App.~\ref{app:rates} for their derivation). 
For instance, the rates describing the tunneling of an electron from the $M$ lead to an empty quasiparticle state $\gamma_j$ is $\Gamma_{M,i\to f} = \Gamma_M \left|u_j\right|^2 n_{\rm F}(E_f-E_i)$, where $i$ and $f$ label the initial and final states with energy $E_i$ and $E_f$. 
$\Gamma_M \left|u_j\right|^2$ is the effective tunneling rate in the wideband limit multiplied by the local projector on the particle-like component of $\gamma_j$.
Since the temperature is small, the Fermi factor $n_{\rm F}$ is a step-like function, activating the tunneling process mainly when $E_f-E_i<0$. 
The opposite is true when we examine the process for extracting one electron from the occupied quasiparticle state $\gamma_j$, meaning that it is activated when $E_f-E_i >0$. Hence, the sign of the current in the $M$ lead changes abruptly when $n_g$ tunes the energy levels of the device across the resonance $E_i=E_f$, as adding/removing electrons from M becomes more favorable.
We show more details in Fig.~\ref{S2}.

In conclusion, a three terminal Coulomb-blockaded device has a highly tunable differential conductance alongside the standard transport suppression inside the Coulomb diamonds.
As shown in panels (c)-(f) of Fig.~\ref{Fig2_v2}, this is a robust feature that requires neither the presence of a zero energy subgap state nor a third lead coupled only to the superconducting continuum, even though the latter is useful to have an extended bias window where the $M$ lead is effectively decoupled from the device.

\section{Double nanowire}\label{sec:DNW}
%
\begin{figure*}[t] \centering
\includegraphics[width=0.9\linewidth]{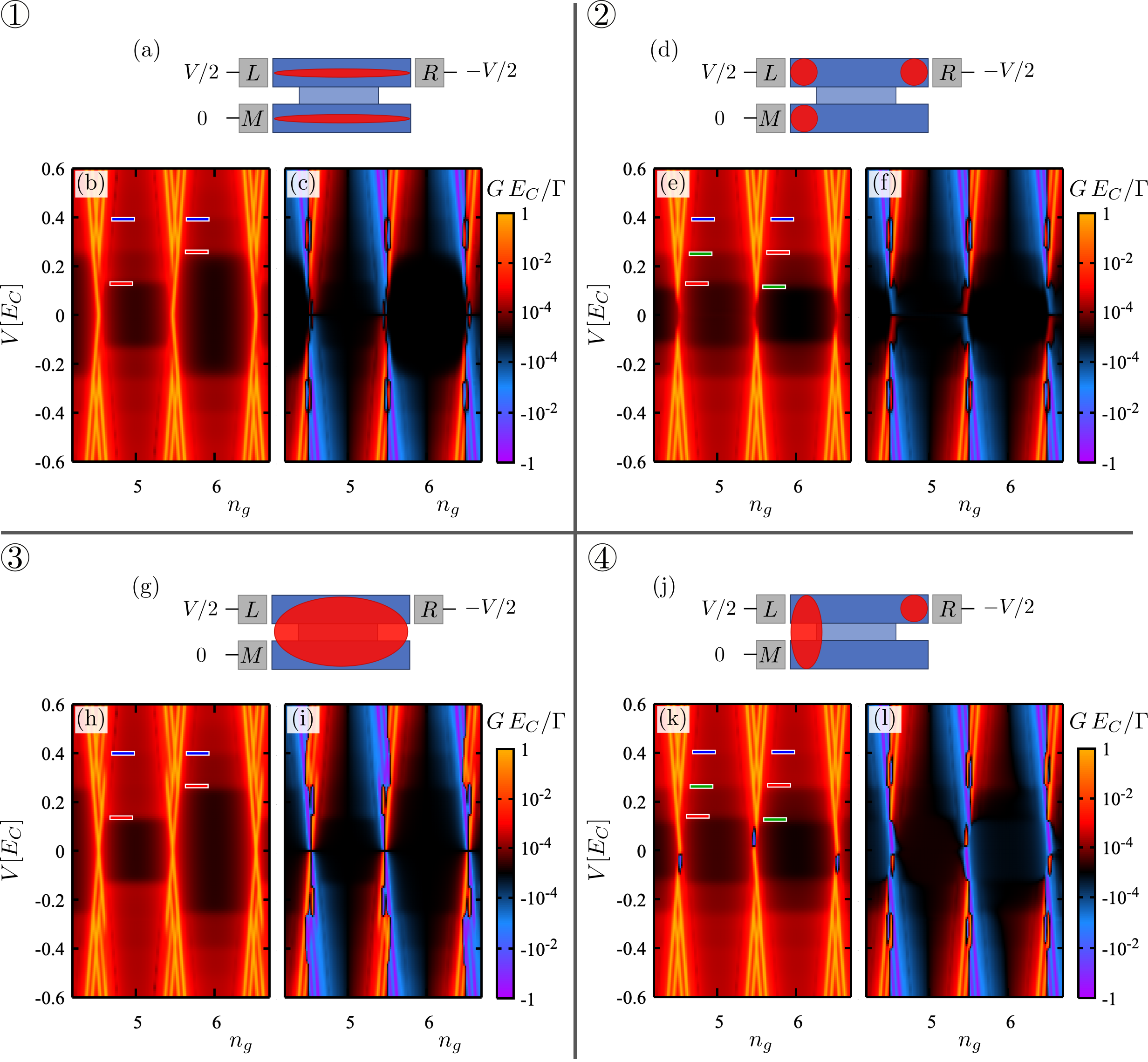}
\caption{Double wire system coupled via a trivial superconductor allowing the exchange of Cooper pairs, so the system has a common charging energy. \raisebox{.5pt}{\textcircled{\raisebox{-.9pt} {$1$}}} One subgap state connects to $L$ and $R$, while another couples to $M$ (a). Panels (b) and (c) are the local  conductance through the lead $L$ and the nonlocal conductance through the grounded $M$ lead, respectively. \raisebox{.5pt}{\textcircled{\raisebox{-.9pt} {$2$}}} Situation for three local states coupled to the different leads, \raisebox{.5pt}{\textcircled{\raisebox{-.9pt} {$3$}}} an extended state coupled to all the leads, and \raisebox{.5pt}{\textcircled{\raisebox{-.9pt} {$4$}}} a local state that couples to $R$, while $L$ and $R$ couple to a different extended state. The parameters are the same as in Fig. \ref{Fig1}. 
}\label{Fig3}
\end{figure*}
%
In this section, we extend the analysis of the nonlocal conductance to double nanowire setups; two parallel semiconductor nanowires are covered by the same floating superconducting island. The common superconductor allows the exchange of Cooper pairs between the wires, making them share a common charging energy, but it does not allow the tunneling of excitations above the gap from one wire to the other. In this context, we therefore introduce two separate states at energy $\Delta$ to model the quasiparticle continua in the two wires. In contrast, we assume that subgap states can delocalize between the two wires.
This setup for the device is inspired by recent experimental achievements~\cite{Kurtossy2021Oct,Vekris_PRB2021,vekris2021asymmetric,Vekris_DNW2022} and the scope of exploring the topological Kondo effect when the system couples to more than two leads~\cite{Beri_PRL2012,Altland_PRL2014,Zazunov_NJP2014}.
We consider that the nanowire ends couple to three different leads, as sketched in panels (a), (d), (g), and (j) of Fig.~\ref{Fig3}. As before, we consider that the $L/R$ leads are symmetrically biased, while $M$ is grounded. A different biasing condition is shown in Fig. \ref{S3} in the appendix.

Regarding the spatial structure of subagap states, we identify four possible situations: \raisebox{.5pt}{\textcircled{\raisebox{-.9pt} {$1$}}} two subgap states that extend along a single nanowire, \raisebox{.5pt}{\textcircled{\raisebox{-.9pt} {$2$}}} a subgap state localized at each lead-device interface, \raisebox{.5pt}{\textcircled{\raisebox{-.9pt} {$3$}}} a common subgap state that couples with all three terminals, \raisebox{.5pt}{\textcircled{\raisebox{-.9pt} {$4$}}} two subgap states localized respectively at the left and right ends of the double nanowire device.
These scenarios are sketched in panels (a), (d), (g), and (j) of Fig.~\ref{Fig3}, while the panels below each sketch show the related local $G_L$ and nonlocal $G_M$ differential conductances. 
In all cases, we consider degenerate subgap states with energy $\varepsilon=0.3 \Delta$. Subgap states with different energies would be easily detected by the different sizes of the Coulomb diamonds.
The differences between the four cases are summarized in Table \ref{Table1}.


\begin{table*}[th]
    \begin{center}
    \begin{tblr}{|[0pt,white]l|[1.2pt,gray]l|[0pt,white]}
    \raisebox{.5pt}{\textcircled{\raisebox{-.9pt} {$1$}}} & \raisebox{.5pt}{\textcircled{\raisebox{-.9pt} {$2$}}}\\
    \hspace{0.75cm}Sequential tunneling absent in $G_M$  & \hspace{0.75cm}Sequential tunneling absent in all terminals  \\
    \hspace{0.75cm}2 cotunneling steps  & \hspace{0.75cm}3 cotunneling steps  \\
    \hline[1.2pt,gray]
    \raisebox{.5pt}{\textcircled{\raisebox{-.9pt} {$3$}}} & \raisebox{.5pt}{\textcircled{\raisebox{-.9pt} {$4$}}}\\
    \hspace{0.75cm}Sequential tunneling present in all terminals & \hspace{0.75cm}Sequential tunneling absent in $G_R$     \\
    \hspace{0.75cm}2 cotunneling steps  & \hspace{0.75cm}3 cotunneling steps  \\
    & \hspace{0.75cm}Negative differential conductance in $G_L$
    \end{tblr}
    \caption{Summary of the main transport features for the four situations shown in Fig. \ref{Fig3}.}
    \label{Table1}
    \end{center}
\end{table*}

Let us start from case \raisebox{.5pt}{\textcircled{\raisebox{-.9pt} {$1$}}}, where each nanowire hosts a single subgap state with support on both ends. In this case, there is no probe at the middle of any of the wires. Therefore, the transport features cannot distinguish between trivial extended states or nonlocal Majorana-like subgap states. The associated local conductance, $G_L = \frac{\ud I_L}{\ud V}$, and the non local one, $G_M = \frac{\ud I_M}{\ud V}$, are reported in Figs.~\ref{Fig3} (b) and (c), respectively. 
Sequential tunneling processes contribute to the current between $L$ and $R$ at low bias, while they are suppressed for the $M$ lead. A single sequential tunneling line is observed in $G_L$ at low bias voltages; in the odd valleys, the stationary distribution $P^{\rm stat}$ has indeed a significant contribution from the configuration where a quasiparticle is frozen in the subgap state in the lower wire. This is due to the imbalance between the $\Gamma_M$ rates for incoming and outgoing particles at small temperatures.  When the lower wire subgap state is occupied, the superconducting island cannot be excited by removing this quasiparticle via sequential tunneling through leads $L$ or $R$, thus suppressing the leading-order transport mechanism in $G_L$. This is evident when comparing the odd valleys in Fig.~\ref{Fig1}(c) and Fig.~\ref{Fig3}(b): in proximity of the charge degeneracy points the sequential tunneling lines moving towards the odd valleys disappear and cotunneling becomes the dominating process at low bias.
Also the weak conductance through the $M$ lead at low bias voltage is due to cotunneling processes, where the device exchanges one electron with either $L$ or $R$, and $M$, keeping the total charge on the superconducting island constant.

In case \raisebox{.5pt}{\textcircled{\raisebox{-.9pt} {$2$}}}, when a trivial subgap state localizes at the interface with each lead, the conductance is suppressed for $|V| < \varepsilon$. This is due to the local support of the subgap states, which cannot directly mediate sequential transport between different leads.
Therefore, transport is dominated by elastic cotunneling at low bias voltages. The nonlocal conductance $G_M$, Fig.~\ref{Fig3}(f), displays feint resonances in the even valleys. The reason behind is that tunneling processes between the superconducting island and the $M$ lead involve (virtual) changes of the number of Cooper pairs in the island. The sign of the current depends on the charge difference between the ground and the lowest excited state.

Case \raisebox{.5pt}{\textcircled{\raisebox{-.9pt} {$3$}}} corresponds to a single subgap state coupled with all three leads. Therefore, the same Coulomb structure is present in the local and the nonlocal conductance, although the latter is an odd function of the bias $V$ and exhibits sign changes when there are level crossings in the many-body spectrum of the superconducting device. The sequential tunneling of electrons dominates transport, leading to strong conductance features at the charge degeneracy points for small bias voltages. Qualitatively, the signal  of the $M$ lead is equivalent to that shown in Fig.~\ref{Fig2_v2}(f) for a single nanowire with an extended bound state.

The last possibility we analyse, case \raisebox{.5pt}{\textcircled{\raisebox{-.9pt} {$4$}}}, corresponds to two degenerate subgap states, localized one in the left part and one in the right part of the device. 
This situation is of particular interest, as it might correspond to a double nanowire geometry where Majorana modes localized at each device ends strongly hybridize due to their small spatial separation. The local conductance $G_L$ displays a parity dependent negative differential conductance (NDC) region at small bias, see Fig.~\ref{Fig3}(k). This is caused by a quasiparticle being trapped in the subgap state coupled to the $R$ lead, blocking the current between the $L$ and $M$ leads. The appearance of the NDC region for either positive or negative bias is due to the electron- or hole-like excitation trapped in the right subgap state. In this regime, sequential transport through the right lead is suppressed, leading to a current reduction (not shown). 
For $|V|>2\varepsilon$, the sequential transport channel to the right lead is open again, therefore making $|G_L|\neq |G_M|$. It also provides a relatively fast relaxation mechanism for the trapped quasiparticles in the right lead and the NDC region disappears.

The four cases show conductance steps inside the Coulomb valleys at finite voltage values. These steps appear when $V$ matches the system excitation energy, opening the inelastic cotunneling channel: exchange of two electrons between the leads and the island, leaving it in an excited state (although keeping the island charge invariant). Depending on whether a bound state couples simultaneously to the $L$ and $R$ leads, cases \raisebox{.5pt}{\textcircled{\raisebox{-.9pt} {$1$}}} and \raisebox{.5pt}{\textcircled{\raisebox{-.9pt} {$3$}}}, or not, cases \raisebox{.5pt}{\textcircled{\raisebox{-.9pt} {$2$}}} and \raisebox{.5pt}{\textcircled{\raisebox{-.9pt} {$4$}}}, the number of cotunneling steps vary from 2 to 3 (colored tics in Fig. \ref{Fig3}). A similar behavior has been described for the single wire situation, see discussion around Fig. \ref{Fig1}.

Finally, we show in Appendix \ref{app:SR} another biasing situation, where $L$ and $M$ are symmetrically biased and $R$ is grounded. We find that local and nonlocal transport can distinguish between the considered four situations also in that case.

\section{conclusions}\label{sec:conclusion}
In this paper, we analysed how the transport features of a multiterminal superconducting device with strong charging energy depend on the number and spatial structure of subgap bound states.
In particular, we investigated the role of the spatial extent of the subgap states, which might not couple to all the leads attached to the device; we showed that the nonlocal differential conductance in multiterminal devices allows for a qualitative characterization of their spatial profiles for biases below the superconducting gap.
We considered a zero bandwidth model to describe the low-energy features of a superconducting floating island consisting of a proximitized single semiconducting nanowire or a pair of parallel nanowires. 
These systems are indeed known to host subgap states that determine the transport properties of the device. 
We focused on a situation where the charging energy $E_{\rm C}$ is the dominant energy scale, 
and we adopted a second-order master equations approach which allows us to characterize both the sequential tunneling of single electrons and the inelastic cotunneling events. 

The sequential tunneling signal, dominating for small leads-device tunneling strength, gives information on the energy and spatial structure of the subgap states mediating transport. When two leads are not coupled by the lowest-energy states, the zero bias conductance is strongly suppressed, leaving only a faint cotunneling feature. In this way, transport can determine whether a state has support on the two ends of the wire. In the same way, additional leads can be added to gain spatial resolution inside the wire. Therefore, the absence of sequential tunneling conductance at low bias is a way to discriminate between trivial extended states and possible topological states with only support at the ends of the wires.
Moreover, the number and voltage of inelastic cotunneling steps allow us to determine the number and the energies of the subgap states. The cotunneling signal is therefore another indicator that can be used to characterize the spatial structure of the subgap states. The described features hold for an arbitrary number of discrete subgap states with energy $|\varepsilon| > T,\ \Gamma$. For $\varepsilon< T,\ \Gamma$, instead, different subgap state configurations may result in the same qualitative transport features, thus hindering their spatial characterization.

Finally, electron transport in multiterminal Coulomb-blockaded devices results in an interesting pattern of peaks with positive and negative differential nonlocal conductance, depending on the induced charge $n_g$ and on the voltage bias $V$. This allows to switch the direction of the current flowing in the grounded lead, or suppress it, without changing the potential difference between the source and the drain but only by tuning the induced charge on the whole superconducting island.  

We presented results based on a perturbative analysis of the transport properties, valid when the temperature is larger than the coupling between the leads and the device. 
Complementary methods are needed to describe the low-temperature and strong-coupling regimes, where electron correlations effects are important, in the 
non-equilibrium situation~\cite{Schmitteckert_PRB2004,Souto_PRB2021,Chung_MPSblockaded}. The master equation approach we presented, however, is less computationally intensive and provides a clear picture of the transport mechanisms as long as non-perturbative effects can be neglected.
.

\section*{aknowledgements}
We acknowledge support from the Danish National Research Foundation, the Danish Council for Independent Research $|$ Natural Sciences, the European Research Council (Grant Agreement No. 856526), the Swedish Research Council, and NanoLund. M.W. and M.B. are supported by the Villum Foundation (Research Grant No. 25310). This project has received funding from the European Union’s Horizon 2020 research and innovation program under the Marie Sklodowska-Curie grant agreement No. 847523 “INTERACTIONS”.

\appendix

\section{Tunneling rates}
\label{Sec::tunnelingRates}
\subsection{Sequential tunneling rates}\label{app:rates}
In this article, we focus on the limit where the charging energy is the largest energy scale ($E_{\rm C}\gg\max\left(|\Delta|,\,|V| \equiv |\mu_{\rm L} - \mu_{\rm R}|\right)$). In this regime, transport is dominated by processes where one electron is exchanged between the superconducting island and the leads. For $\Gamma$ much smaller than any other energy scale, sequential tunneling processes yield the most prominent features in the conductance, providing the usual Coulomb diamond structure.
The sequential rates are given by
    \begin{align}
    \begin{split}
         	\Gamma^{\rm{seq}}_{\nu,\left|N,N_{\rm C},\textbf{n}\right\rangle\to\left|N+1,N_{\rm C},\textbf{n'}\right\rangle}&=\Gamma_\nu \left|u_j\right|^2 n_{\rm F}(E_f-E_i-\mu_\nu)\,\\
	        \Gamma^{\rm{seq}}_{\nu,\left|N,N_{\rm C},\textbf{n}\right\rangle\to\left|N-1,N_{\rm C},\textbf{n'}\right\rangle}&=\Gamma_\nu \left|u_j\right|^2 n_{\rm F}(\mu_\nu+E_f-E_i)\,\\
	        \Gamma^{\rm{seq}}_{\nu,\left|N,N_{\rm C},\textbf{n}\right\rangle\to\left|N+1,N_{\rm C}+1,\textbf{n'}\right\rangle}&=\Gamma_\nu \left|v_j\right|^2 n_{\rm F}(E_f-E_i-\mu_\nu)\,\\
         	\Gamma^{\rm{seq}}_{\nu,\left|N,N_{\rm C},\textbf{n}\right\rangle\to\left|N-1,N_{\rm C}-1,\textbf{n'}\right\rangle}&=\Gamma_\nu \left|v_j\right|^2 n_{\rm F}(\mu_\nu+E_f-E_i)\,,
    \end{split}
    \end{align}
where $n_{\rm F}$ is the Fermi-Dirac distribution function and $\mu_\nu$ is the chemical potential of the lead $\nu=L,R,M$.
Here, we have used the wideband approximation, where the tunneling rates are energy independent and $\Gamma_\nu=2\pi\rho_F |t_\nu|^2$, with the lead density of states at the Fermi level $\rho_F$. These rates induce transitions between charge states differing by one electron, denoted through the vectors $\textbf{n}$ and $\textbf{n}'$ accounting for the fermionic occupation of the island states.

\subsection{Cotunneling rates} \label{app:cot}
Inside the Coulomb-blockaded region, inelastic cotunneling creates a series of steps in the differential conductance, where the voltage bias corresponds to energy differences between different states of the superconducting island. These are processes where one electron is transferred between two leads, leaving the island in an excited state. We also consider the elastic cotunneling that, instead, describes processes in which the island energy is conserved, and gives rise to the conductance background inside the diamond. However, we disregard local and crossed Andreev processes, where the island charge changes by 2e, as they are suppressed by the strong charging energy in the system. The corresponding rates are given by Eq. \eqref{rates_def}, where 
    \begin{equation}
        \begin{split}
    	F_{\nu,\nu'}(\omega_1) & \equiv \left|\left\langle f\left| \mathcal{T}\right|i\right\rangle\right|^2 = \\
    	& \left| \sum_{m_1,\nu}\frac{\left\langle f\left| H_{\rm T}(\nu) \right|m_1\right\rangle\left\langle m_1\left| H_{\rm T}({\nu'}) \right|i\right\rangle}{E_{m_1}-E_i-\omega_1} \right. \\
    	&\left. +\sum_{m_2,\nu} \frac{\left\langle f\left| H_{\rm T}({\nu'}) \right|m_2\right\rangle\left\langle m_2\left| H_{\rm T}(\nu) \right|i\right\rangle}{E_{m_2}-E_f+\omega_1}\right|^2\,.
        \end{split}
        \label{Kernel_2order}
    \end{equation}
Here, $H_{\rm T}(\nu)$ describes the tunneling between the lead $\nu$ and the island, $\nu'$ denotes a  lead different from $\nu$, and  $m_{1,2}$ are virtual intermediate states. To derive this expression, we have imposed energy conservation, which leads to a function dependent on the energy of the tunneling electron from/to one of the leads, $\omega_{1}$.
The cotunneling rate can be written as
    \begin{equation}
        \begin{split}
        	\Gamma^{\rm{cot}}_{\nu,\nu',i\to f}=2\pi\int d\omega_1\,& F_{\nu,\nu'}(\omega_1)\, n_{\rm F}(\omega_1-\mu_{\rm L})\,\\
        	\times\, & n_{\rm F}(\mu_{\rm R}+E_f-E_i-\omega_1)\,.
        	\label{cot_rate}
        \end{split}
    \end{equation}
This expression for the cotunneling rates is divergent when the denominator in Eq. \eqref{Kernel_2order} vanishes. To avoid the divergent behaviour, we regularize the divergences as explained in Ref. \cite{Koller2010}. The resulting integral can be formally solved analytically~\cite{Koch2004}, which leads to a complicated expression involving special functions. We find that for $T/E_{\rm C}\gtrsim10^{-3}$, it is computationally more efficient to expand the Fermi distribution function into a sum of complex Matsubara-Ozaki frequencies~\cite{Ozaki2007}
\begin{equation}
	n_{\rm F}(\omega-\mu)=\sum_\alpha r_\alpha\frac{1}{\omega-\mu+i\beta_\alpha}\,,
\end{equation}
where $\beta_\alpha$ and $r_\alpha$ are the approximated Matsubara frequencies and residues, respectively.
Finally, Eq.~\eqref{cot_rate} can be evaluated using the residue theorem, yielding to a rather compact expression
    \begin{equation}
        \begin{split}
    	&\Gamma^{\rm{cot}}_{\nu,\nu',i\to f} =  \,8\pi\, \mbox{Im}\sum_{\alpha=0}^\infty r_\alpha \left\{ n_{\rm F}(\mu_{\rm R}+E_f-E_i-\mu_{\rm L}+i\beta_\alpha) \right. \\ 
    	 &\left. \times \left[F_{\nu,\nu'}(\mu_{\rm L}-i\beta_\alpha) - F_{\nu,\nu'}(\mu_{\rm R}+E_f-E_i+i\beta_\alpha)\right] \right\}\,.
        \end{split}
    \end{equation}
This sum can be truncated at $\alpha\approx 100$ Matsubara-Ozaki frequencies for the parameters used in the calculations.

\section{Supporting results}\label{app:SR}

In this appendix, we present additional results to support the chosen modelling of the considered devices and describe some details of the transport features presented in the main text. In particular we first focus on the processes describing the transport of the three-terminal single-nanowire devices, which determine the transport across the lead $M$; then we address the comparison between different modelling of the quasiparticle states above the superconducting gap; finally, we consider a different choice of the voltage drop across the double-nanowire devices.

\begin{figure}[t] \centering
\includegraphics[width=\columnwidth]{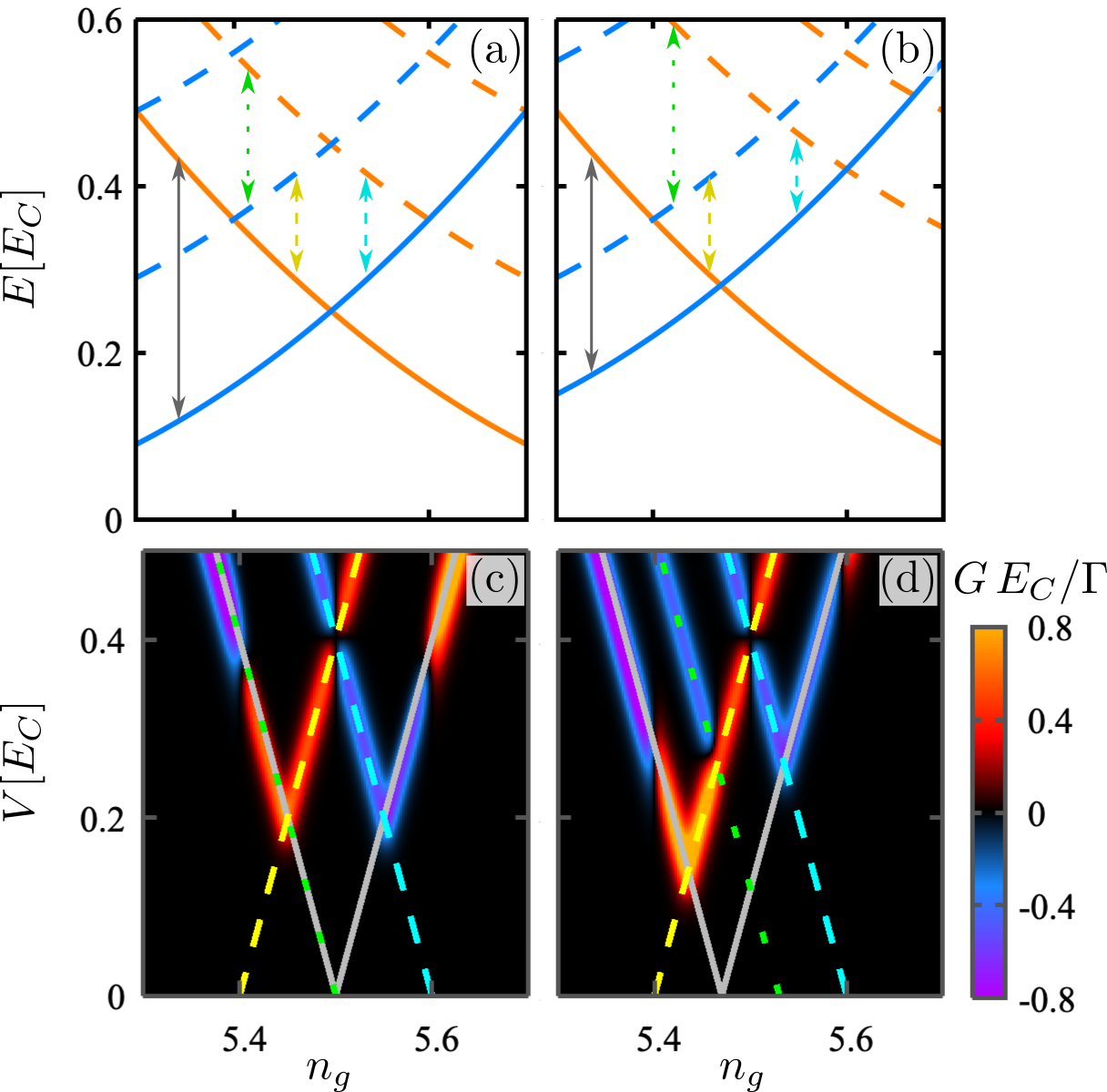}
\caption{(a) Many-body eigenergies of the island for a spinless state at zero energy ($\Delta=0.2E_{\rm C}$). (b) Many-body energies for a subgap state at finite energy, $\varepsilon = 0.3 \Delta$. (c) Differential conductance through a middle lead that does not couple to the subgap state [see the sketch in Fig. \ref{Fig2_v2} (a)]. Here we consider only the sequential tunneling contribution to the current. 
Panel (d) shows the corresponding results for $\varepsilon=0.3\Delta$. The lines in the lower panels correspond to the excitation threshold denoted by the arrows in the upper ones.
In panels (a) and (b), sequential tunneling via the $M$ lead connects only states with different parity and charge occupation in the states at $\Delta$. These processes always involve a relaxation from a higher energy state to a lower energy one, thus explaining the changes in sign of the differential conductance when two parabolas cross. The remaining parameters are the same as in Fig. \ref{Fig1}
}\label{S2}
\end{figure}

\subsection{Features of the current across the lead $M$ in the three-terminal nanowire}

To understand the three-terminal transport features in a Coulomb-blockaded superconducting device, it is useful to compare the conductance dependence on the voltage bias and on the induced charge with the low-energy many-body spectrum. 
In Fig.~\ref{S2}, we report the eigenvalue structure and the nonlocal conductance $G_M=\frac{\ud I_M}{\ud V}$, close to the charge degeneracy point,  for the device sketched in Fig.~\ref{Fig2_v2}(a): a single Coulomb-blockaded nanowire hosting a subgap state that does not couple with the grounded lead $M$.
The lowest energies of states with odd and even parity are represented by the continuous blue and orange lines, respectively, in Fig.~\ref{S2}(a)-(b). These states are connected via sequential tunneling involving a particle transfer to or from the subgap state, which does not contribute to the current flowing through the $M$ lead. Hence $G_M$ is suppressed  at low bias, as can be seen in Figs.~\ref{S2}(c) and (d).
The current in the $M$ lead is activated only when $V$ is large enough to populate excited states (via the $L$ or the $R$ leads) that can then relax to a lower energy state through a tunneling event between the superconducting continuum and the lead $M$. These processes correspond to transitions between a dashed line (i.e., states with excited quasiparticles in the superconducting continuum) and a continuous line with different parity (colors) in Figs.~\ref{S2} (a) and (b).
These events are highlighted by the oblique lines in panels (c) and (d), representing energy thresholds for populating states that can contribute to the current through $M$. The sign of the current, and hence of the conductance, changes at the crossing between energy levels with different parity.
This phenomenology is independent from the energy $\varepsilon$ of the subgap states, whether it is zero, as in panels (a) and (c), or finite, as in panels (b) and (d). The latter situation only has a slightly richer structure of the Coulomb diamonds due to the energy difference between the first excited states in the even and odd sectors.

\begin{figure*} \centering
\includegraphics[width=0.9\linewidth]{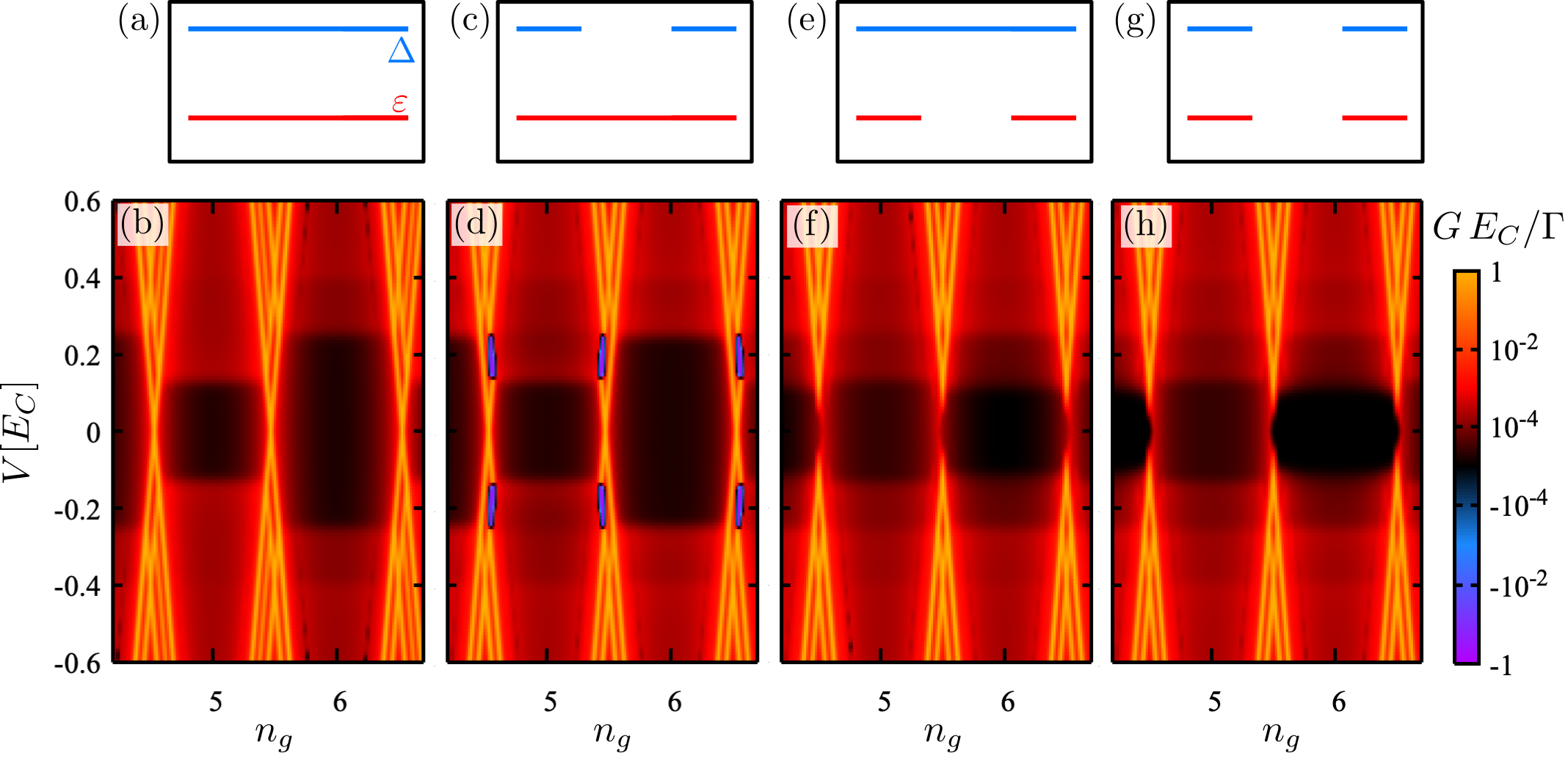}
\caption{Two-terminal conductance for different subgap and above gap states situations. Panels (a) and (c) show schematically the situation where a bound state extends from left to right (red), while the state at higher energy can be either extended (a) or local (c). Panels (b) and (d) show the corresponding local conductance. Panels (e) and (g) show the situation where the island hosts local subgap states, whose corresponding conductance is shown in (f) and (h).
Notice the NDC region appearing in panel (d): they correspond to a particle blocked in one of the two state at energy $\Delta$ and unable to tunnel out from the device on the other side, thus suppressing transport.
Below the energy threshold required to populate the superconducting continuum ($|V|=2\Delta$), transport is dominated by the properties of the subgap state alone: the low-bias peaks of panels (b) and (d) are almost identical and so are those of panels (f) and (h).
}\label{S1}
\end{figure*}

Another interesting effect shown in Fig.~\ref{Fig2_v2} is the sign mismatch between the sequential tunneling and the cotunneling contribution close to the lower edge of the Coulomb diamonds. This can be seen at $V=\Delta$, where cotunneling (blurred signal) gives rise to negative conductance while sequential tunneling (sharp lines) contributes with positive conductance when $n_g $ approaches $5.5$ from below.
Both can be understood by considering the processes mediated by tunneling through the $M$ lead for specific values of $n_g$ and $V$.
When $V>0$ and $n_g \to 5.5$, the lowest energy excitation that can relax through the central lead is the odd parity state with a quasiparticle in the superconducting continuum 
that couples with an incoming electron to create a Cooper pair. This process is thus associated with a positive (ingoing) current from lead $M$.
Instead, the lowest energy inelastic cotunneling step in the odd valley corresponds to the excitation of a quasiparticle from the subgap state to the continuum, mediated by the virtual occupation of the even parity state with no quasiparticle present.
In this second-order process, the only tunneling event through the $M$ lead is the destruction of a Cooper pair into the high-energy quasiparticle and an {\em outgoing} electron, which carries a negative particle current and, thus, negative conductance.
Similar arguments explain the sign change between sequential tunneling and cotunneling in other regions of the Coulomb diamonds.

\subsection{One- vs two-state approximation to model the Bogoliubov quasiparticle continuum}

Concerning the modelling of the quasiparticle continuum above the gap, in Fig.~\ref{S1} we compare the two-terminal conductance for a  nanowire with a single quasiparticle state at energy $\Delta$ coupled with both leads and with two degenerate quasiparticle states at the same energy with finite supports on each end, as sketched in the upper panels. The nanowire hosts one or two subgap states at energy $\varepsilon$. 
As long as the voltage bias is smaller than $\Delta-\varepsilon$, transport is dominated by the properties of the subgap state and the spatial structure of the superconducting continuum does not affect the qualitative features in conductance of the device.
When the bias is larger and the states at energy $\Delta$ become accessible via sequential tunneling events, the ``broken'' continuum can trap a quasiparticle in one of the two edges of the device, suppressing transport and inducing a negative differential conductance region in the Coulomb diamonds.
Notice, however, how this NDC is qualitatively different from that appearing in all data presented in the main text, as it is a sequential tunneling feature appearing in the local conductance of two-terminal devices.

\begin{figure*} \centering
\includegraphics[width=0.9\linewidth]{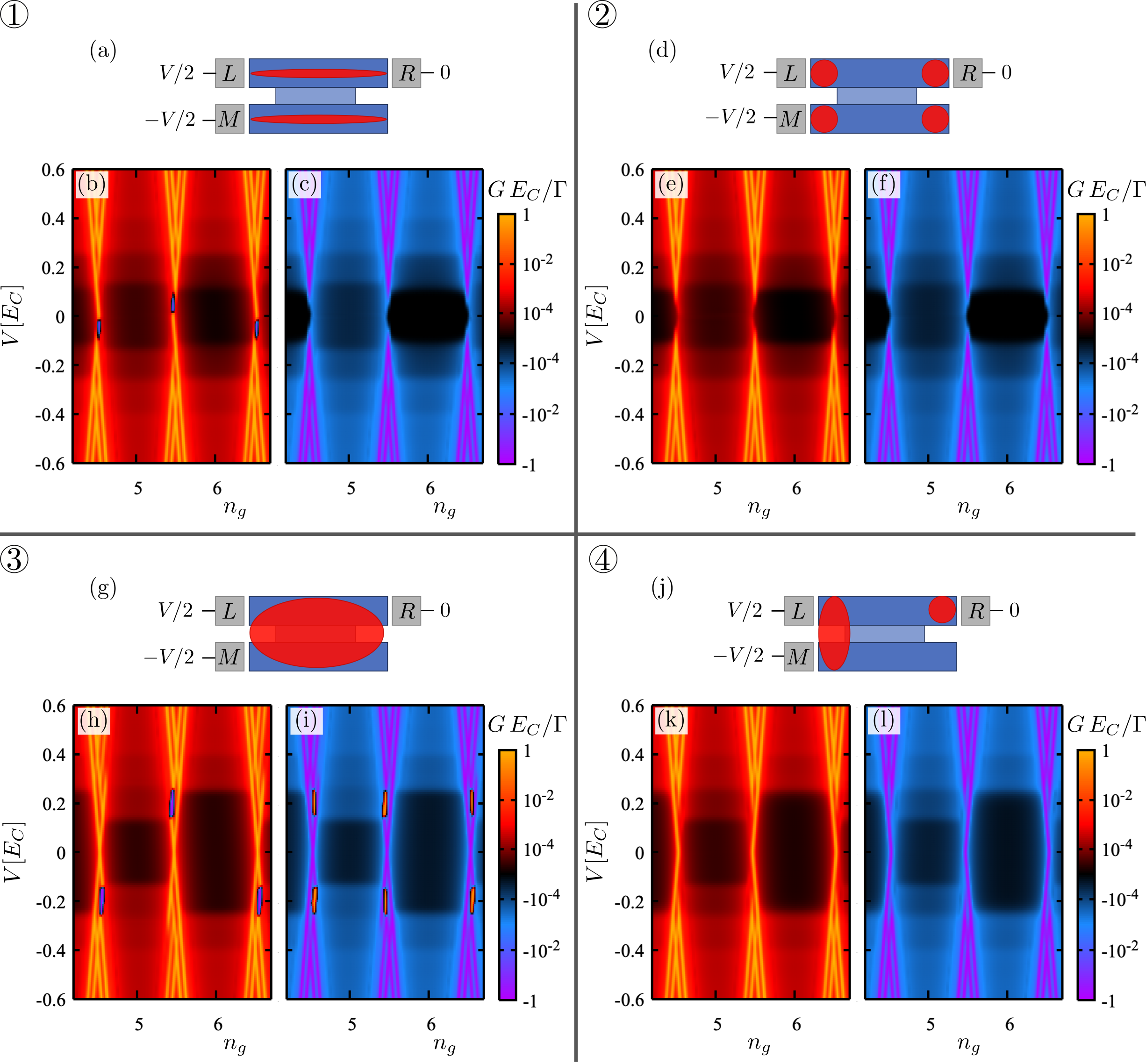}
\caption{Upper left panels: sketch of the double wire system. \raisebox{.5pt}{\textcircled{\raisebox{-.9pt} {$1$}}} one bound state connects $L$ and $R$, while $M$ connects to another one (a). Panels (b) and (c) are the conductance through the grounded $M$ lead. \raisebox{.5pt}{\textcircled{\raisebox{-.9pt} {$2$}}} three local states coupled to the different leads, \raisebox{.5pt}{\textcircled{\raisebox{-.9pt} {$3$}}} an extended state coupled to all the leads, and \raisebox{.5pt}{\textcircled{\raisebox{-.9pt} {$4$}}} a local state that couples to $R$, while $L$ and $R$ couple to another extended state. The parameters are the same as in Fig. \ref{Fig1}.
The conductance we show for lead $M$ is $G_M=\frac{\ud I_M}{\ud V_R}$, hence it is always negative, with the exception of some small regions.
}\label{S3}
\end{figure*}

\subsection{Example of the multiterminal transport with different voltage drops}

Finally, in Fig.~\ref{S3} we present the conductance results for the double nanowire geometry with a different bias choice with respect to the data shown in Fig. \ref{Fig3} in the main text: here we bias symmetrically leads $L$ and $M$, while lead $R$ is left grounded.
The main difference with respect to Fig.~\ref{Fig3} is that now the biased leads are not connected by a single superconducting continuum at energy $\Delta$.
In particular, notice the different natures of the NDC regions appearing in panel (b) and (h): in the former, it is due to a quasiparticle trapped in the subgap state connected to lead $M$ and, indeed, it is not reflected in $G_M$. In the latter, it corresponds to a quasiparticle trapped in one of the two superconducting continua, in the upper or in the lower nanowire.
This second case is similar to the data presented in Fig.\ref{S1}(d), where the NDC is due to the broken superconducting continuum on a single nanowire, although the conductance in Fig.~\ref{S3}(h) is not symmetric in $V$ because of the presence of the third terminal.


%

\end{document}